\documentclass[aps,pra,reprint,superscriptaddress,amsmath,amssymb]{revtex4-2}

\usepackage{graphicx}
\usepackage{braket}
\usepackage{hyperref}

\begin{document}

\title{Insecurity of Measurement-Device-Independent Quantum Key Distribution}

\author{Konstantin Zaitsev}
\email{Zaitsev20k@gmail.com}
\affiliation{Independent Researcher, Vigo, Spain}

\author{Polina Acheva}
\affiliation{Independent Researcher, Vigo, Spain}

\begin{abstract}
The security of practical quantum key distribution (QKD) systems is fundamentally constrained by vulnerabilities of single-photon detectors. Measurement-device-independent quantum key distribution (MDI-QKD) was proposed to remove this limitation by allowing all measurements to be performed by a completely untrusted party, under the assumption that the measurement node can be treated as adversarial but does not compromise the security guarantees of the protocol. Here we show that this assumption is insufficient under realistic adversarial control of the measurement device. We present an attack in which an adversary exploits active control of the measurement node (Charlie) to obtain significant information about the secret key. The attack enables recovery of up to 70\% of the sifted key while introducing only 5.6\% quantum bit error rate. Unlike previously reported attacks targeting specific implementations of MDI-QKD, our results demonstrate a limitation of the standard security model underlying the protocol. These findings indicate that additional constraints on the measurement-device independence assumption, or refined security analyses incorporating stronger adversarial capabilities, are required to ensure the security of MDI-QKD in realistic scenarios.
\end{abstract}

\maketitle

\section{Introduction}\label{sec:intro}

Quantum key distribution (QKD) is often presented as a unique cryptographic technology whose security can be derived directly from the laws of physics~\cite{bennett_quantum_1984, ekert_quantum_1991}. As a consequence, discussions of QKD security have traditionally focused on the validity of security proofs~\cite{mayers_unconditional_2001, lo_unconditional_1999, shor_simple_2000}, while practical attacks have been interpreted as failures of implementation rather than failures of theory~\cite{brassard_limitations_2000, gisin_trojan-horse_2006, makarov_effects_2006, lydersen_hacking_2010}.

This distinction shaped the development of quantum hacking as a research field. Over the past two decades, numerous attacks have demonstrated that practical QKD systems may exhibit exploitable side channels arising from imperfect state preparation~\cite{fung_phase-remapping_2007, huang_laser_2019}, detector vulnerabilities~\cite{makarov_effects_2006, zhao_quantum_2008, lydersen_hacking_2010, gerhardt_full-field_2011}, optical components~\cite{li_attacking_2011, sun_passive_2011}, and calibration procedures~\cite{jain_device_2011}. These results significantly improved the understanding of implementation security, yet they rarely challenged the underlying theoretical models themselves.

As a result, theoretical security analyses have largely remained outside the scope of adversarial scrutiny traditionally applied to practical systems. Security proofs are commonly treated as definitive descriptions of protocol security, while attacks are expected to reveal implementation weaknesses rather than limitations of the underlying model.

However, any security proof is necessarily derived within a particular set of assumptions. The validity of the resulting security claim therefore depends not only on the correctness of the mathematical derivation, but also on the completeness and physical relevance of the assumptions on which the derivation is based. In computational cryptography, security claims are routinely framed in terms of explicit assumptions and known limitations of the underlying model. This naturally raises the question of which assumptions are required for the security claims associated with modern QKD protocols.

Measurement-device-independent quantum key distribution (MDI-QKD), originally proposed as a protocol based on two-photon interference~\cite{lo_mdi_2012}, represents a particularly interesting case. Originally introduced as a countermeasure against detector-side attacks, MDI-QKD was designed to remove the need to trust single-photon detectors, which historically represented one of the most vulnerable components of practical QKD systems.

In this work, we present a concrete attack against MDI-QKD. The attack relies exclusively on linear optical elements, including polarization transformations, and requires photon-number-resolving detectors in place of the threshold detectors commonly assumed in the standard protocol. Unlike conventional quantum hacking attacks, it does not exploit imperfections of a particular implementation. Instead, it targets the protocol itself. Beyond the specific attack presented here, our objective is to examine why such an attack becomes possible within a protocol that is generally regarded as secure.

\section{MDI-QKD protocol description}\label{sec:protocol}

\begin{figure}[t]
\centering
\includegraphics[width=\columnwidth]{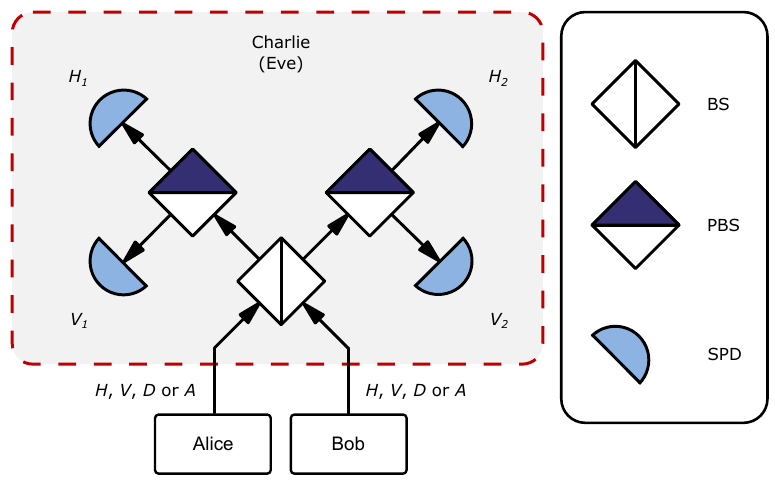}
\caption{Schematic diagram of the MDI-QKD setup. Alice and Bob, the legitimate users, each send one of four polarization states, $H$, $V$, $D$, or $A$, to the untrusted relay Charlie. Charlie performs a Bell-state measurement and publicly announces either the $\psi_-$ outcome (coincidence between $H_1$ and $V_2$ or between $H_2$ and $V_1$) or the $\psi_+$ outcome (coincidence between $H_1$ and $V_1$ or between $H_2$ and $V_2$). BS, 50:50 beam splitter; PBS, polarizing beam splitter; SPD, single-photon detector.}
\label{fig:mdi_scheme}
\end{figure}

For conceptual clarity, we consider the original MDI-QKD protocol introduced in Ref.~\cite{lo_mdi_2012} using ideal single-photon states, although practical implementations typically employ weak coherent pulses together with decoy-state analysis. We further restrict the discussion to polarization encoding, noting that the same protocol can also be implemented using alternative encodings such as phase encoding. These simplifications do not affect the central argument presented in this work.

In the considered protocol (see Fig.~\ref{fig:mdi_scheme}), Alice and Bob independently prepare single-photon states in one of two mutually unbiased bases. The computational $Z$ basis corresponds to horizontal and vertical polarization states $(H,V)$, while the complementary $X$ basis corresponds to diagonal polarization states $(D,A)$. Each user encodes a classical bit value into the prepared quantum state, for example by associating $H$ and $D$ with logical ``1'', and $V$ and $A$ with logical ``0''.

The prepared states are then transmitted to a third party, traditionally called Charlie, who performs a Bell-state measurement. Importantly, Charlie is treated as an untrusted party and may even be fully controlled by an adversary. Using linear optics, the measurement setup can unambiguously distinguish only two Bell states, typically denoted $\Psi^{-}$ and $\Psi^{+}$. Whenever such a projection is successfully detected, Charlie publicly announces the corresponding measurement outcome.

After Charlie's announcement, Alice and Bob publicly reveal only the bases used for state preparation while keeping the encoded bit values secret. Events corresponding to incompatible preparation bases are discarded. For the remaining events, Alice uses her original bit value as the reference raw-key bit. Bob then applies a deterministic correction rule based on both the announced Bell-state outcome and the common preparation basis. In the $Z$ basis, Bob flips his bit for both $\Psi^{-}$ and $\Psi^{+}$ outcomes. In the $X$ basis, Bob flips his bit for $\Psi^{-}$ outcomes and keeps it unchanged for $\Psi^{+}$ outcomes. After this correction step, Alice and Bob obtain identical raw-key bits.

The protocol then proceeds through standard QKD post-processing stages, including error-rate estimation, error correction, and privacy amplification. A randomly selected subset of the raw key is publicly compared in order to estimate the quantum bit error rate (QBER). If the observed error exceeds the security threshold allowed by the protocol, secret key generation is aborted. Otherwise, the remaining information potentially available to an adversary is reduced through privacy amplification, yielding the final secret key.

Each successful Bell-state announcement implicitly corresponds to a three-class classification problem. For a $\Psi^{-}$ announcement, the input state combinations are divided into error events $(HH, VV, DD, AA)$, which should never produce a $\Psi^{-}$ outcome, bit-1 events $(HV, DA)$, and bit-0 events $(VH, AD)$. Similarly, for a $\Psi^{+}$ announcement, the corresponding classes are error events $(HH, VV, DA, AD)$, bit-1 events $(HV, DD)$, and bit-0 events $(VH, AA)$~\cite{russell_bell_state_analyzer_mdi_qkd}.

According to Zeilinger's foundational principle, an elementary quantum system carries one bit of information~\cite{zeilinger_1999_foundational}. Consequently, a measurement performed on two photons can reveal at most two bits of information about the incoming state. Under this constraint, no measurement strategy can simultaneously distinguish the two valid key classes while perfectly separating them from the corresponding error class. Any attempt to obtain more reliable information about the sifted-key bit inevitably increases the probability of misclassifying error events, whereas a strategy optimized to reject error events necessarily sacrifices information about the key.

\section{Attack description}\label{sec:attack}

\subsection{Adversarial exploitation of the protocol}

The argument above relies on an implicit assumption: Charlie's decision is based exclusively on measurements performed on the two photons received from Alice and Bob. Under this assumption, the available information is fundamentally limited by the two-photon system itself.

However, the MDI-QKD protocol imposes no restriction on the internal operation of an untrusted relay. Charlie is defined only by his input-output behavior and may be implemented in any physically allowed manner.

In this work, we relax the implicit two-photon constraint by considering an adversarial implementation in which Eve extends the physical system with two auxiliary photons prepared in known polarization states.

\begin{figure}[t]
\centering
\includegraphics[width=\columnwidth]{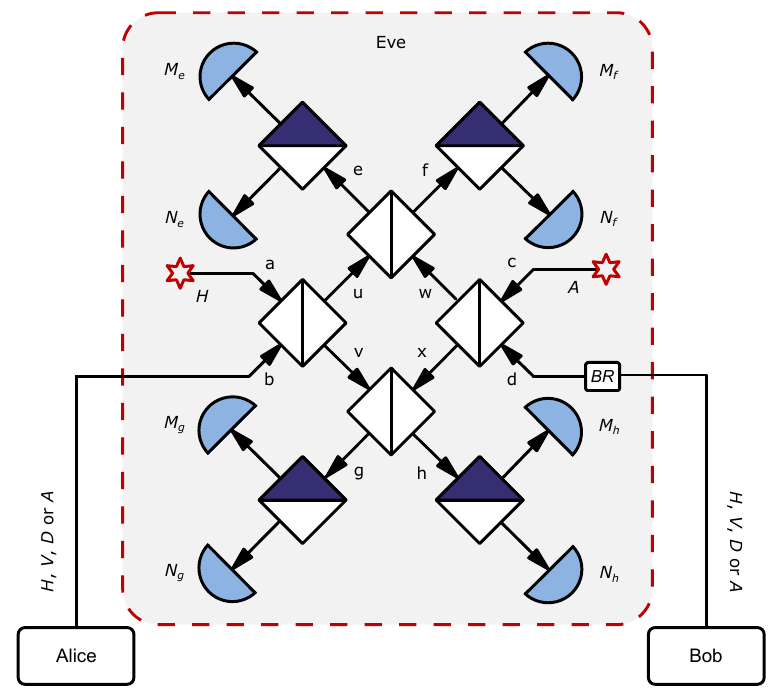}
\caption{Schematic of the auxiliary-photon relay implementing a post-selection attack through linear-optical processing and photon-number-resolving detection of extended input states. BR (bit rotation) denotes the conditional transformation applied to Bob's bit depending on Eve's selected $\Psi^{-}$ or $\Psi^{+}$ emulation branch (see Steps~1--2 in the main text for details).}
\label{fig:relay_scheme}
\end{figure}

Eve replaces the honest Bell-state measurement relay with a modified linear-optical measurement setup, referred to here as the auxiliary-photon relay (see Fig.~\ref{fig:relay_scheme}). The device is implemented using standard linear-optical components, including beam splitters, polarization beam splitters, and photon-number-resolving detectors. It realizes a post-selection-based measurement strategy by processing incoming optical signals through additional routing and conditioning stages, while still producing valid relay announcements consistent with the MDI-QKD interface.

The introduction of this additional internal structure modifies the effective measurement resolution of the system. In particular, the space of distinguishable detection signatures depends on the physical implementation of the relay. In the idealized photon-number-resolving description of the honest Charlie receiver, this space reduces to eight elementary signatures, corresponding to four single-detector click events $(2H_1, 2H_2, 2V_1, 2V_2)$, two $\Psi^{-}$-type coincidence patterns $(H_1V_2, V_1H_2)$, and two $\Psi^{+}$-type coincidence patterns $(H_1V_1, H_2V_2)$.

In contrast, the auxiliary-photon relay admits a significantly richer set of detection signatures due to the enlarged optical Hilbert space. In the idealized model considered here, this results in 258 distinguishable detection signatures when photon-number resolution is taken into account.

The MDI-QKD protocol, however, does not require the relay to announce all observed detection events. It only constrains the external interface to valid Bell-state announcements. This freedom enables Eve to implement a post-selection strategy, where only a subset of detection signatures is declared as $\Psi^{-}$ and $\Psi^{+}$ outcomes, while all other events are discarded. We therefore refer to this construction as a \emph{post-selection attack}.

In particular, Eve only selects two four-element classes of detection signatures. The first class,
\[
2M_h + 2N_h,\quad
2M_g + 2N_g,\quad
2M_f + 2N_f,\quad
2M_e + 2N_e,
\]
is denoted as \emph{Same}. The second class,
\[
2M_f + 2N_h,\quad
2M_h + 2N_f,\quad
2M_e + 2N_g,\quad
2M_g + 2N_e,
\]
is denoted as \emph{Opposite}. Here, $2M_e$ denotes a two-photon detection event in mode $M_e$, i.e., the corresponding detector registers exactly two photons. The remaining terms are defined analogously.

It is worth noting that these structures are visually reminiscent of the coincidence patterns associated with $\Psi^{+}$ and $\Psi^{-}$ outcomes in ideal Bell-state measurements. However, this correspondence is purely structural: the Same/Opposite partition does not imply that $\Psi^{+}$ or $\Psi^{-}$ announcements are generated exclusively from one class. In particular, both classes can contribute to either Bell-state outcome depending on the internal processing strategy of the relay.

An additional distinction between the auxiliary-photon relay and an honest Charlie receiver lies in the choice of measurement basis used for extracting information from the incoming optical signals. In the honest implementation, the measurement is fixed by the Bell-state projection structure and is not designed to optimize information about the encoded bits.

\begin{figure}[t]
\centering
\includegraphics[width=0.75\columnwidth]{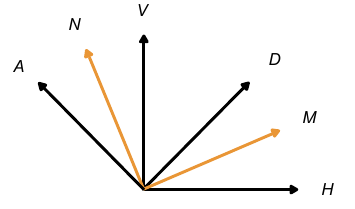}
\caption{Illustration of the intermediate measurement basis (interbasis) employed in the auxiliary-photon relay for extracting bit information in the post-selection attack.}
\label{fig:interbasis}
\end{figure}

In contrast, since Eve is not constrained by knowledge of the preparation basis used by Alice and Bob, her goal is solely to extract information about the encoded bit value. For this purpose, the auxiliary-photon relay performs measurements in an intermediate basis defined by the orthonormal states
\[
|M\rangle = \frac{|X_1\rangle + |Z_1\rangle}{\sqrt{2 + \sqrt{2}}}, \qquad
|N\rangle = \frac{|X_0\rangle + |Z_0\rangle}{\sqrt{2 + \sqrt{2}}},
\]
with $|X_1\rangle = (|Z_1\rangle + |Z_0\rangle)/\sqrt{2}$ and $|X_0\rangle = (|Z_0\rangle - |Z_1\rangle)/\sqrt{2}$, which is unbiased with respect to both the $Z$ and $X$ preparation bases. In the polarization encoding considered here, this intermediate basis corresponds to a rotation of the polarization states and is represented by the intermediate basis shown in Fig.~\ref{fig:interbasis}, given by
\[
\begin{aligned}
H &= cM - sN, \qquad & V &= sM + cN, \\
D &= cM + sN, \qquad & A &= sM - cN,
\end{aligned}
\]
where $c = \cos(\pi/8)$ and $s = \sin(\pi/8)$.

This choice of basis allows Eve to extract partial information about the encoded bit without requiring knowledge of the preparation basis in advance, and complements the post-selection stage of the attack by refining the internal assignment of detection signatures to bit values.

To isolate the analysis from implementation-specific imperfections, we assume idealized conditions including unit detection efficiency, lossless channels, and noiseless optical components. The relay is further assumed to employ photon-number-resolving detectors capable of distinguishing different multiphoton events.

Finally, Alice and Bob are assumed to prepare ideal single-photon states in the four polarization states $H$, $V$, $D$, and $A$, each chosen with equal probability. Eve is assumed to have full control over the auxiliary photon source used in the relay.

The attack itself proceeds as follows.

\textbf{Step 1.} Eve selects the Bell-state projection that she intends to emulate, namely either the $\psi_-$ or the $\psi_+$ outcome.

\textbf{Step 2.} Depending on the Bell-state measurement outcome selected in Step~1, Eve applies one of the following transformation strategies. The polarization state of the photon sent by Alice is left unchanged in both cases; Eve only modifies the polarization state of the photon sent by Bob.

\emph{Emulation of the $\psi_-$ outcome.} Eve applies the transformation
\[
H \rightarrow D,\qquad
D \rightarrow H,\qquad
V \rightarrow A,\qquad
A \rightarrow V
\]
to the photon sent by Bob. This transformation exchanges the rectilinear and diagonal bases while preserving the encoded bit value. The incoming polarization is first decomposed in the $22.5^\circ$ basis, a relative phase shift of $\pi$ is applied to one arm, and the two paths are then recombined interferometrically.

\emph{Emulation of the $\psi_+$ outcome.} Eve applies the transformation
\[
H \rightarrow D,\qquad
D \rightarrow V,\qquad
V \rightarrow A,\qquad
A \rightarrow H
\]
to the photon sent by Bob. This operation corresponds to a $45^\circ$ rotation of the polarization state on the Bloch sphere.

\textbf{Step 3.} Regardless of whether Eve emulates the $\psi_-$ or the $\psi_+$ outcome, the subsequent stage of the attack is identical. Eve injects the transformed photons from Alice and Bob into input modes $b$ and $d$ of the auxiliary-photon relay. At the same time, she prepares two auxiliary single photons under her full control and injects them into input modes $a$ and $c$. In the attack considered here, these auxiliary photons are prepared in the states $H$ and $A$, respectively. Eve then performs the measurement implemented by the relay.

\textbf{Step 4.} Eve accepts a detection event if and only if it belongs to either the \emph{Same} or \emph{Opposite} class defined above. All other detection signatures are discarded. For accepted events, Eve publicly announces a successful Bell-state measurement corresponding to the outcome selected in Step~1.

\textbf{Step 5.} After Alice and Bob publicly announce their preparation bases, events corresponding to incompatible bases are discarded according to the standard sifting procedure. Eve applies the same sifting rule and retains only the surviving events. The bit assignment rule depends only on the \emph{Same} and \emph{Opposite} classes defined above and is independent of the announced Bell-state label. Events belonging to the \emph{Same} class are interpreted as bit value $0$ in the $H/V$ basis and bit value $1$ in the $D/A$ basis. Events belonging to the \emph{Opposite} class are interpreted as bit value $1$ in the $H/V$ basis and bit value $0$ in the $D/A$ basis.

\textbf{Step 6.} Finally, Eve applies a classical post-processing strategy analogous to that used by Alice and Bob. She does not know the sifted key deterministically. Instead, for each retained event she assigns a candidate bit value inferred from the publicly announced Bell-state measurement outcome and the observed detection pattern.

The \emph{post-selection attack} introduces a QBER of approximately $5.6\%$. At the same time, Eve obtains information about approximately $70\%$ of the sifted key bits. Although this error rate exceeds typical values reported in experimental MDI-QKD demonstrations, it remains within the operational regime where positive secret-key generation is generally still possible. A more detailed discussion is provided in Sec.~\ref{sec:discussion}.

\subsection{Attack design principles}

\begin{table}[t]
\caption{Transformation of Alice's and Bob's input states under the preparation steps used in the $\Psi^{-}$ and $\Psi^{+}$ branches of the attack. The labels indicate whether the corresponding original input state belongs to the error class, Alice's bit-$1$ class, or Alice's bit-$0$ class.}
\label{tab:attack_state_classes}
\begin{ruledtabular}
\begin{tabular}{ccc}
Initial state & $\Psi^{-}$ preparation & $\Psi^{+}$ preparation \\
\hline
$HH$ & $HD$ (error) & $HD$ (error) \\
$VV$ & $VA$ (error) & $VA$ (error) \\
$HV$ & $HA$ (bit $1$) & $HA$ (bit $1$) \\
$VH$ & $VD$ (bit $0$) & $VD$ (bit $0$) \\
$DD$ & $DH$ (error) & $DV$ (bit $1$) \\
$AA$ & $AV$ (error) & $AH$ (bit $0$) \\
$DA$ & $DV$ (bit $1$) & $DH$ (error) \\
$AD$ & $AH$ (bit $0$) & $AV$ (error) \\
\end{tabular}
\end{ruledtabular}
\end{table}

The motivation behind the state transformations introduced in Step~2 can be understood by examining how input states are mapped under the two Bell-state measurement branches.

For the $\Psi^{-}$ branch, the states $HH$, $VV$, $DD$, and $AA$ cannot yield a valid $\ket{\Psi^-}$ outcome in the honest protocol, and therefore any such event announced as successful necessarily contributes to the error class. The remaining accepted states split into two groups according to Alice's bit assignment: $HV$ and $DA$ correspond to bit $1$, while $VH$ and $AD$ correspond to bit $0$.

For the $\Psi^{+}$ branch, the structure is rearranged: $HH$, $VV$, $DA$, and $AD$ form the error class, while $HV$ and $DD$ map to bit $1$, and $VH$ and $AA$ map to bit $0$.

As summarized in Table~\ref{tab:attack_state_classes}, both branches are reduced by the transformations in Step~2 to a consistent classification structure over transformed detection events. This allows the same optical processing stage to handle both Bell-state outcomes, while the distinction between $\Psi^{-}$ and $\Psi^{+}$ is delegated to the announced measurement result rather than the internal optical evolution.

\subsection{Quantum-optical attack model}

The attack is analyzed using the standard Fock-state formalism of quantum optics. Optical states are represented by photon creation operators acting on the vacuum state. In particular, an $n$-photon state occupying optical mode $a$ is written as
\begin{equation}
\ket{n_a}
=
\frac{\left(\hat a_a^\dagger\right)^n}{\sqrt{n!}}
\ket{0}.
\label{eq:fock_state}
\end{equation}

Beam splitters are described by the corresponding unitary transformations of the creation operators. For a lossless $50{:}50$ beam splitter with input modes $a$ and $b$ and output modes $c$ and $d$,
\begin{align}
\hat a_{a,\mu}^{\dagger}
&\rightarrow
\frac{1}{\sqrt{2}}
\left(
\hat a_{c,\mu}^{\dagger}
+
i\hat a_{d,\mu}^{\dagger}
\right),
\nonumber\\
\hat a_{b,\mu}^{\dagger}
&\rightarrow
\frac{1}{\sqrt{2}}
\left(
i\hat a_{c,\mu}^{\dagger}
+
\hat a_{d,\mu}^{\dagger}
\right),
\label{eq:beam_splitter}
\end{align}
where $\mu$ denotes the polarization mode.

\subsection{State evolution through the auxiliary-photon relay}

The attack analysis begins by expressing all polarization states in the $M/N$ basis introduced above,
\[
\begin{aligned}
H &= cM - sN, \\
V &= sM + cN, \\
D &= cM + sN, \\
A &= sM - cN,
\end{aligned}
\]
where $c=\cos(\pi/8)$ and $s=\sin(\pi/8)$.

In the attack considered here, Eve injects an $H$-polarized photon into spatial mode $a$ and an $A$-polarized photon into spatial mode $c$. The photons originating from Alice and Bob, after the polarization transformations described in Step~2, occupy spatial modes $b$ and $d$, respectively. Their polarization states are represented in the general form $\alpha M+\beta N$ and $\gamma M+\delta N$. This parametrization allows all transformed BB84 states to be treated within a single expression.

The corresponding four-photon input state is
\begin{equation}
\begin{aligned}
\ket{\Psi_{\mathrm{in}}}
=
&\left(
c \hat a_{a,M}^\dagger
-
s \hat a_{a,N}^\dagger
\right)
\left(
\alpha \hat a_{b,M}^\dagger
+
\beta \hat a_{b,N}^\dagger
\right)
\\
&\times
\left(
s \hat a_{c,M}^\dagger
-
c \hat a_{c,N}^\dagger
\right)
\left(
\gamma \hat a_{d,M}^\dagger
+
\delta \hat a_{d,N}^\dagger
\right)
\ket{0}.
\end{aligned}
\label{eq:psi_in}
\end{equation}

The state evolution through the auxiliary-photon relay is obtained by successive application of the beam-splitter transformations introduced in Eq.~(\ref{eq:beam_splitter}). The first beam-splitter layer mixes spatial modes $(a,b)$ and $(c,d)$, while the second layer mixes modes $(u,w)$ and $(v,x)$. Applying these transformations to $\ket{\Psi_{\mathrm{in}}}$ yields the four-photon state immediately before polarization-resolved detection,
\begin{align}
\ket{\Psi_{\mathrm{out}}}
=
\frac{1}{16}
&
\Big[
c
\left(
\hat a_{e,M}^\dagger
+
i\hat a_{f,M}^\dagger
+
i\hat a_{g,M}^\dagger
-
\hat a_{h,M}^\dagger
\right)
\nonumber\\
&
-
s
\left(
\hat a_{e,N}^\dagger
+
i\hat a_{f,N}^\dagger
+
i\hat a_{g,N}^\dagger
-
\hat a_{h,N}^\dagger
\right)
\Big]
\nonumber\\
&\times
\Big[
\alpha
\left(
i\hat a_{e,M}^\dagger
-
\hat a_{f,M}^\dagger
+
\hat a_{g,M}^\dagger
+
i\hat a_{h,M}^\dagger
\right)
\nonumber\\
&
+
\beta
\left(
i\hat a_{e,N}^\dagger
-
\hat a_{f,N}^\dagger
+
\hat a_{g,N}^\dagger
+
i\hat a_{h,N}^\dagger
\right)
\Big]
\nonumber\\
&\times
\Big[
s
\left(
i\hat a_{e,M}^\dagger
+
\hat a_{f,M}^\dagger
-
\hat a_{g,M}^\dagger
+
i\hat a_{h,M}^\dagger
\right)
\nonumber\\
&
-
c
\left(
i\hat a_{e,N}^\dagger
+
\hat a_{f,N}^\dagger
-
\hat a_{g,N}^\dagger
+
i\hat a_{h,N}^\dagger
\right)
\Big]
\nonumber\\
&\times
\Big[
\gamma
\left(
-\hat a_{e,M}^\dagger
+
i\hat a_{f,M}^\dagger
+
i\hat a_{g,M}^\dagger
+
\hat a_{h,M}^\dagger
\right)
\nonumber\\
&
+
\delta
\left(
-\hat a_{e,N}^\dagger
+
i\hat a_{f,N}^\dagger
+
i\hat a_{g,N}^\dagger
+
\hat a_{h,N}^\dagger
\right)
\Big]
\ket{0}.
\label{eq:psi_out}
\end{align}

The measurement strategy was constructed by considering individual detector click patterns and identifying the input-state classes that maximize their discrimination power. As an illustrative example, consider the detector state
\[
2M_h + 2N_h.
\]

To calculate the corresponding probability amplitude, we project $\ket{\Psi_{\mathrm{out}}}$ onto the four-photon Fock state with two photons in mode $M_h$ and two photons in mode $N_h$. Equivalently, in Eq.~(\ref{eq:psi_out}) we retain only the terms containing the creation operators $\hat a^{\dagger}_{h,M}$ and $\hat a^{\dagger}_{h,N}$. For this projection, the relevant part of $\ket{\Psi_{\mathrm{out}}}$ is
\[
\begin{aligned}
\ket{\Psi_{\mathrm{out}}} \rightarrow \frac{1}{16}
&\left(
- c \hat a^{\dagger}_{h,M} + s \hat a^{\dagger}_{h,N}
\right)
\left(
i\alpha \hat a^{\dagger}_{h,M} + i\beta \hat a^{\dagger}_{h,N}
\right)
\\
&\times
\left(
i s \hat a^{\dagger}_{h,M} - i c \hat a^{\dagger}_{h,N}
\right)
\left(
\gamma \hat a^{\dagger}_{h,M} + \delta \hat a^{\dagger}_{h,N}
\right)
\ket{0}.
\end{aligned}
\]

Next we expand the resulting expression and simplify it by collecting identical operator products and retaining only the non-vanishing contributions. The projection coefficient acquires an additional factor of $2$, since
\[
\left(\hat a_{h,M}^{\dagger}\right)^2
\left(\hat a_{h,N}^{\dagger}\right)^2
\ket{0}
=
2\ket{2M_h,2N_h},
\]
according to the Fock-state normalization in Eq.~(\ref{eq:fock_state}).

All detector states within the \emph{Same} class lead to identical contributions. The corresponding probability amplitude is therefore common to the entire class and is given by
\begin{equation}
A_{\mathrm{same}}
=
-\frac{1}{8}
\left[
\alpha\delta
+
\beta\gamma
-
sc
\left(
\alpha\gamma
+
\beta\delta
\right)
\right].
\label{eq:amp_same}
\end{equation}

All detector states belonging to the \emph{Opposite} class also yield the same probability amplitude,
\begin{equation}
A_{\mathrm{opposite}}
=
-\frac{1}{8}
\left[
\alpha\delta
+
\beta\gamma
+
sc
\left(
\alpha\gamma
+
\beta\delta
\right)
\right].
\label{eq:amp_opposite}
\end{equation}

Thus, the eight retained detector states are described by only two distinct probability amplitudes. All detector states within a given class therefore produce identical detection probabilities. These eight detector states exhibit the strongest correlation with Eve's three transformed state classes and form the basis of the measurement strategy considered in this work.

For the transformed states considered in this attack, the coefficients $\alpha,\beta,\gamma,\delta$ are completely determined by the polarization states prepared by Alice and Bob. Since both parties choose among the four BB84 polarization states, each coefficient is equal to either $\pm c$ or $\pm s$. As an example, for the transformed input state $HD$,
\[
\alpha=c,\qquad
\beta=-s,\qquad
\gamma=c,\qquad
\delta=s.
\]
Substitution into Eqs.~(\ref{eq:amp_same}) and (\ref{eq:amp_opposite}) gives
\[
A_{\mathrm{same}}^{(HD)}
=
-\frac{sc(c^2-s^2)}{8},
\qquad
A_{\mathrm{opposite}}^{(HD)}
=
\frac{sc(c^2-s^2)}{8}.
\]

The corresponding probabilities are obtained by taking the modulus square of the amplitudes. The resulting amplitudes are summarized in Table~\ref{tab:projection}.

\begin{table}[t]
\caption{Classification of Alice's and rotated Bob's input states and the corresponding projection amplitudes.}
\label{tab:projection}
\begin{ruledtabular}
\begin{tabular}{cccc}
Input state & Class & $A_{\mathrm{same}}$ & $A_{\mathrm{opposite}}$ \\
\hline
\begin{tabular}{@{}c@{}}
$HD \quad DH$ \\
$VA \quad AV$
\end{tabular}
& Error &
$\pm\dfrac{sc(c^2-s^2)}{8}$ &
$\pm\dfrac{sc(c^2-s^2)}{8}$ \\
\begin{tabular}{@{}c@{}}
$HA$ \\
$AH$
\end{tabular}
&
\begin{tabular}{@{}c@{}}
Bit 1 \\
Bit 0
\end{tabular}
&
$\dfrac{1+2(sc)^2}{8}$ &
$\dfrac{1-2(sc)^2}{8}$ \\
\begin{tabular}{@{}c@{}}
$VD$ \\
$DV$
\end{tabular}
&
\begin{tabular}{@{}c@{}}
Bit 0 \\
Bit 1
\end{tabular}
&
$\dfrac{1-2(sc)^2}{8}$ &
$\dfrac{1+2(sc)^2}{8}$ \\
\end{tabular}
\end{ruledtabular}
\end{table}

As can be seen from Table~\ref{tab:projection}, the detector states $A_{\mathrm{same}}$ and $A_{\mathrm{opposite}}$ are observed predominantly for the transformed states belonging to the bit-$1$ and bit-$0$ classes. At first sight, this may appear problematic, since both classes contribute to the same detector signature.

However, this ambiguity is resolved during the sifting stage. After the measurement, Alice and Bob publicly announce the preparation basis used for each signal. Since Eve's transformations modify only Bob's polarization state while leaving Alice's state unchanged, the first symbol in each transformed state directly identifies Alice's basis choice. Consequently, once the basis information is revealed, the ambiguity between the state pairs $(HA,AH)$ and $(DV,VD)$ is removed. Although the states within each pair produce identical detector statistics, they correspond to different bit values and originate from different preparation bases. This can be seen directly from the first symbol of each transformed state, since Alice's state is left unchanged by the transformations in Step~2.

This basis-dependent disambiguation is implemented in Step~5, where Eve uses the publicly announced basis information together with her detector outcome to assign the corresponding bit value.

The probabilities associated with the retained detector states are obtained as the squared moduli of the corresponding probability amplitudes. They take only three distinct values,
\begin{align*}
P_1 &= \frac{\left[sc(c^2-s^2)\right]^2}{64} = 0.0009765625,\\
P_2 &= \frac{\left[1-2(sc)^2\right]^2}{64} = 0.0087890625,\\
P_3 &= \frac{\left[1+2(sc)^2\right]^2}{64} = 0.0244140625.
\end{align*}

The total probability associated with this detector pattern is therefore
\[
P_{\mathrm{tot}} = 4P_1 + 2P_3 + 2P_2 = 0.0703125.
\]
The probability that Eve's announcement results in an error is
\[
P_{\mathrm{err}} = \frac{4P_1}{P_{\mathrm{tot}}} = 0.0556.
\]
Similarly, the probability that Eve correctly infers the corresponding sifted-key bit is
\[
P_{\mathrm{Eve}} = \frac{2P_3}{P_{\mathrm{tot}}} = 0.6944.
\]

Thus, for this detector pattern, Eve correctly identifies the sifted-key bit with probability $69.4\%$ while introducing a quantum bit error rate of only $5.6\%$.

The same procedure is applied to all detector patterns retained by Eve. The resulting probabilities are then combined according to the frequencies of the corresponding detector events, yielding the overall quantum bit error rate introduced by the attack and Eve's average information about the sifted key.

\section{Limitations of the Attack}\label{sec:discussion}

In this section, we discuss the limitations of the proposed attack and possible directions for the development of countermeasures.

\subsection{QBER threshold}

The proposed attack introduces a QBER of approximately $5.6\%$ in the raw key. To assess whether this value is large or small, it is useful to compare it with the tolerable error rates of the protocol.

In the BB84 protocol, attacks are often characterized by the induced QBER and compared with the well-known asymptotic threshold of approximately $11\%$. For MDI-QKD, however, no universal threshold exists, and the tolerable error rate depends strongly on the particular implementation and the assumptions used in the security proof. This is partly because MDI-QKD was developed within the framework of mature composable security theory and incorporates various corrections accounting for realistic imperfections and finite-size effects.

As an example, let us consider the finite-key analysis presented in Ref.~\cite{curty_finite-key_2014}. According to Eq.~(1) of that work, the length of the final secret key is given by
\begin{equation}
l = \underline{n}_{11}^Z \left[ 1 - H_2(e_{11}^X) \right]
- \mathrm{leak}_{EC}
- \Delta_{FK},
\label{eq:key_length}
\end{equation}
where $\underline{n}_{11}^Z$ denotes the lower bound on the number of single-photon events contributing to the sifted key, $e_{11}^X$ is the estimated phase error rate, $\mathrm{leak}_{EC}$ represents the amount of information disclosed during error correction, and $\Delta_{FK}$ accounts for finite-size effects.

To gain intuition, let us compare Eq.~(\ref{eq:key_length}) with the simplified model considered in this work. First, we assume ideal single-photon sources. In this case, all sifted-key bits originate from single-photon events, and it is convenient to normalize the secret-key length per successful single-photon detection event. Consequently, the prefactor $\underline{n}_{11}^Z$ can be omitted. Second, we consider the asymptotic limit of infinitely long keys and neglect finite-size effects, i.e., $\Delta_{FK}=0$. Finally, since $\mathrm{leak}_{EC}$ appears only during classical post-processing, we postpone its consideration and focus on the raw correlations produced by the attack.

Under these assumptions, Eq.~(\ref{eq:key_length}) reduces to
\begin{equation}
l = 1-H_2(e_{11}^X).
\label{eq:key_length2}
\end{equation}
Thus, in this idealized model the secret key length is determined solely by the phase error rate. Assuming the asymptotic symmetric case, $e_{11}^X=e_{11}^Z=e$, Eq.~(\ref{eq:key_length2}) yields the familiar threshold $H_2(e)=1/2$, corresponding to $e\approx 11\%$. Therefore, a QBER below $11\%$ would traditionally be interpreted as allowing secret key extraction.

This threshold originates from the information-theoretic security analysis of BB84 developed by Shor and Preskill~\cite{shor_simple_2000}, where security is established against the most general coherent attacks. In this framework, Eve interacts coherently with the transmitted quantum states by attaching ancillas, stores them in a quantum memory, and performs an optimal joint measurement only after the public discussion stage. Since Bob measures the disturbed quantum states, Eve's information gain is intrinsically linked to the disturbance introduced into the transmitted signals, giving rise to the well-known information--disturbance trade-off and, ultimately, to the asymptotic $11\%$ threshold.

The security proofs of MDI-QKD inherit this information-theoretic framework. They assume that Eve performs coherent attacks on the quantum signals propagating from Alice and Bob to the relay, while the Bell-state measurement itself is represented only through its prescribed input-output behavior. In contrast, the attack proposed here exploits the fact that the relay is physically controlled by Eve. She has direct access to the photons entering the measurement device and can engineer the measurement itself rather than merely interacting coherently with the transmitted states. Consequently, the attack is not constrained by the information--disturbance trade-off underlying the coherent-attack paradigm and can produce a QBER of only $5.6\%$.

The simplified model described above should be regarded only as an upper bound corresponding to idealized asymptotic conditions. In practice, realistic MDI-QKD systems operate under much more stringent constraints. In particular, the finite-key analysis of Ref.~\cite{curty_finite-key_2014} demonstrates that positive secret-key generation is achieved only for QBER values of the order of $0.5\%$--$2.5\%$ (see Fig.~4 of Ref.~\cite{curty_finite-key_2014}). Thus, the gap between the asymptotic threshold and practical operation is largely determined by finite-size effects and other non-idealities included in the complete security analysis.

If the error statistics induced by the proposed attack are incorporated into the parameter estimation of practical MDI-QKD implementations, the resulting estimate of the phase error rate enters a regime that lies outside the positivity region of the finite-key key-rate expression. Consequently, the secret key length becomes non-positive, i.e., $l \leq 0$, and no secret key can be extracted within the standard MDI-QKD security framework.

\subsection{Trusted Charlie}

The proposed attack exploits a specific aspect of the MDI-QKD security model: although the measurement node is explicitly declared untrusted, the underlying security analysis continues to rely on a coherent-attack framework originally formulated for trusted-receiver scenarios.

Within the MDI-QKD security model, Charlie is indeed treated as a completely untrusted party, in the sense that the measurement device may be operated by an adversary. Consequently, the attack presented in this work remains fully consistent with the intended threat model of MDI-QKD.

One may then ask whether the same reasoning remains valid if the role of Charlie is modified, i.e., if the measurement node is promoted from an untrusted relay to a trusted network component. In this case, Eve can no longer directly manipulate the Bell-state measurement process. Instead, her strategy must be modified.

A possible approach is for Eve to intercept the quantum signals sent by Alice and Bob and perform the same measurements using the auxiliary-photon relay. Upon obtaining a favorable measurement outcome, Eve does not publicly announce it. Instead, she attempts to reproduce the corresponding event at Charlie's station. This could potentially be achieved using detector-control techniques, detector-efficiency mismatch attacks, or by preparing and transmitting an appropriate quantum state designed to trigger the desired Bell-state announcement. The feasibility of such attacks depends on the specific implementation and lies beyond the scope of the present work.

The important observation is that once Charlie is considered a trusted device, the security analysis must again include assumptions about the behavior of Charlie's detectors. In particular, the system can no longer be regarded as immune to detector-side-channel attacks. Additional detector characterization and security analysis become necessary.

This observation is noteworthy because eliminating detector vulnerabilities was one of the primary motivations behind the development of MDI-QKD. Therefore, promoting Charlie to a trusted node may mitigate the attack described in this work, but it does so at the cost of partially abandoning one of the original security goals of the protocol.

\subsection{Rejected events}

So far, we have considered only those events that contribute to the sifted key, since events corresponding to incompatible basis choices are normally discarded. However, for attacks such as the one proposed here, the rejected data may itself contain nontrivial statistical structure and therefore could, in principle, be used as an additional monitoring channel without introducing extra measurements.

Indeed, the attack induces a non-uniform structure within the rejected events. After grouping detector patterns according to the same public Bell-state announcement, the discarded data exhibit three distinct probability classes with values $0.017578125$, $0.001953125$, and $0.0078125$. In principle, this structure could be used as a statistical signature of the attack and therefore as a potential monitoring tool.

However, several factors limit the usefulness of such a countermeasure. First, these statistics correspond only to the specific realization of the attack considered here and do not reflect an intrinsic property of the protocol. In particular, the probabilities $0.017578125$ and $0.001953125$ are exchanged between the \emph{Same} and \emph{Opposite} classes. Moreover, an eavesdropper aware of a monitoring strategy based on these statistics could easily symmetrize the attack. For instance, Eve could randomly alternate between the original $M$-$N$ basis and a basis rotated by $90^\circ$. This would induce an additional polarization flip $H\leftrightarrow V$, $D\leftrightarrow A$ in the second stage of the attack and replace the internal $H$-$A$ encoding inside the auxiliary-photon relay with $H$-$D$. Averaging over the original attack and its $90^\circ$-rotated variant makes the statistics of the rejected events invariant under this transformation and removes the asymmetry.

Therefore, although monitoring discarded events may reveal signatures of a particular implementation, these signatures are not fundamental and can be eliminated by simple modifications of the eavesdropping strategy. Consequently, post-selection analysis of rejected events does not provide a robust countermeasure against the proposed attack.

\subsection{Key rate}

Another limitation of the proposed attack is the reduction of the secret-key generation rate. As shown above, the attack preserves only a subset of the Bell-state measurement outcomes. Consequently, the probability of a successful Bell-state announcement is reduced from approximately $50\%$ in the ideal honest case to about $7\%$, corresponding to a decrease by roughly a factor of seven.

In principle, such a reduction could be detected by monitoring the raw-key generation rate. However, this observation should be interpreted with caution. The attack assumes that Eve is not constrained by the losses of the legitimate quantum channel. Therefore, part or all of this reduction could be compensated if Eve is able to replace the original channel with one having lower attenuation.

For a standard optical fiber with an attenuation coefficient of approximately $0.2~\mathrm{dB/km}$, a sevenfold reduction in the announcement rate corresponds to about $8.5~\mathrm{dB}$. In the symmetric MDI-QKD configuration, where both Alice and Bob are connected to Charlie by channels of equal length, this loss penalty corresponds to approximately $42~\mathrm{km}$ of total fiber length, or about $21~\mathrm{km}$ for each user.

This value can be compared with practical long-distance QKD networks based on trusted nodes and one-way QKD. For example, in the Beijing--Shanghai quantum backbone network, neighboring trusted nodes are typically separated by distances of the order of $63.8~\mathrm{km}$~\cite{chen_space_ground_2021}. Thus, the reduction introduced by the proposed attack effectively decreases the maximum transmission distance of MDI-QKD by a distance corresponding to nearly one half of the spacing between neighboring nodes in an existing one-way QKD network.

Consequently, although the attack does not necessarily prevent secret-key generation, it significantly reduces its ability to distribute keys over long distances without trusted intermediate stations.

\section{Conclusion}\label{sec:conclusion}

In this work, we have demonstrated an attack against the MDI-QKD protocol that does not depend on a particular implementation and can provide Eve with information about approximately $70\%$ of the sifted key while introducing a QBER of only approximately $5.6\%$.

The attack is based on the introduction of auxiliary photons with known polarization states into an apparatus emulating the Bell-state measurement. This increases the number of possible projection outcomes and allows Eve to perform state discrimination with a significantly higher probability of success. By selectively announcing only favorable measurement outcomes, Eve can obtain substantial information about the sifted key while maintaining a relatively low error rate.

We further argue that the attack presented here is unlikely to be optimal. In principle, the same approach could be extended by introducing a larger number of auxiliary photons or by exploiting detectors with higher photon-number resolution, thereby increasing the number of distinguishable measurement outcomes. Such extensions may further improve Eve's information gain while maintaining a comparable disturbance level. Moreover, the present attack treats each Bell-state measurement independently. It remains an open question whether collective or coherent generalizations acting jointly on multiple rounds could provide Eve with even more information while preserving a similarly low QBER.

These results show that the security proof of MDI-QKD does not fully account for all attack strategies included in its own adversarial model.

\begin{acknowledgments}
The authors declare no external funding for this work.
\end{acknowledgments}

\bibliography{sn-bibliography}

@inproceedings{bennett_quantum_1984,
  author = {Bennett, Charles H. and Brassard, Gilles},
  title = {Quantum Cryptography: Public Key Distribution and Coin Tossing},
  booktitle = {Proceedings of the International Conference on Computers, Systems and Signal Processing},
  address = {Bangalore, India},
  publisher = {IEEE Press},
  year = {1984},
  pages = {175--179}
}

@article{ekert_quantum_1991,
author = {Ekert, Artur K.},
title = {Quantum Cryptography Based on Bell's Theorem},
journal = {Physical Review Letters},
year = {1991},
volume = {67},
number = {6},
pages = {661--663},
doi = {10.1103/PhysRevLett.67.661}
}

@article{mayers_unconditional_2001,
author = {Mayers, Dominic},
title = {Unconditional Security in Quantum Cryptography},
journal = {Journal of the ACM},
year = {2001},
volume = {48},
number = {3},
pages = {351--406},
doi = {10.1145/382780.382781}
}

@article{lo_unconditional_1999,
author = {Lo, Hoi-Kwong and Chau, Hoi Fung},
title = {Unconditional Security of Quantum Key Distribution over Arbitrarily Long Distances},
journal = {Science},
year = {1999},
volume = {283},
number = {5410},
pages = {2050--2056},
doi = {10.1126/science.283.5410.2050}
}

@article{shor_simple_2000,
author = {Shor, Peter W. and Preskill, John},
title = {Simple Proof of Security of the BB84 Quantum Key Distribution Protocol},
journal = {Physical Review Letters},
year = {2000},
volume = {85},
number = {2},
pages = {441--444},
doi = {10.1103/PhysRevLett.85.441}
}

@article{brassard_limitations_2000,
author = {Brassard, Gilles and L{\"u}tkenhaus, Norbert and Mor, Tal and Sanders, Barry C.},
title = {Limitations on Practical Quantum Cryptography},
journal = {Physical Review Letters},
year = {2000},
volume = {85},
number = {6},
pages = {1330--1333},
doi = {10.1103/PhysRevLett.85.1330}
}

@article{gisin_trojan-horse_2006,
author = {Gisin, Nicolas and Fasel, Stefan and Kraus, Barbara and Zbinden, Hugo and Ribordy, Gr{\'e}goire},
title = {Trojan-Horse Attacks on Quantum-Key-Distribution Systems},
journal = {Physical Review A},
year = {2006},
volume = {73},
number = {2},
pages = {022320},
doi = {10.1103/PhysRevA.73.022320}
}

@article{makarov_effects_2006,
author = {Makarov, Vadim and Anisimov, Andrey and Skaar, Johannes},
title = {Effects of Detector Efficiency Mismatch on Security of Quantum Cryptosystems},
journal = {Physical Review A},
year = {2006},
volume = {74},
number = {2},
pages = {022313},
doi = {10.1103/PhysRevA.74.022313},
note = {Erratum: Physical Review A 78, 019905 (2008)}
}

@article{lydersen_hacking_2010,
author = {Lydersen, Lars and Wiechers, Carlos and Wittmann, Christoffer and Elser, Dominik and Skaar, Johannes and Makarov, Vadim},
title = {Hacking Commercial Quantum Cryptography Systems by Tailored Bright Illumination},
journal = {Nature Photonics},
year = {2010},
volume = {4},
pages = {686--689},
doi = {10.1038/nphoton.2010.214}
}

@article{zhao_quantum_2008,
author = {Zhao, Yi and Fung, Chi-Hang Fred and Qi, Bing and Chen, Christine and Lo, Hoi-Kwong},
title = {Quantum Hacking: Experimental Demonstration of Time-Shift Attack against Practical Quantum-Key-Distribution Systems},
journal = {Physical Review A},
year = {2008},
volume = {78},
number = {4},
pages = {042333},
doi = {10.1103/PhysRevA.78.042333}
}

@article{gerhardt_full-field_2011,
author = {Gerhardt, Ilja and Liu, Qin and Lamas-Linares, Antia and Skaar, Johannes and Kurtsiefer, Christian and Makarov, Vadim},
title = {Full-Field Implementation of a Perfect Eavesdropper on a Quantum Cryptography System},
journal = {Nature Communications},
year = {2011},
volume = {2},
pages = {349},
doi = {10.1038/ncomms1348}
}

@article{jain_device_2011,
author = {Jain, Nitin and Wittmann, Christoffer and Lydersen, Lars and Wiechers, Carlos and Elser, Dominique and Marquardt, Christoph and Makarov, Vadim and Leuchs, Gerd},
title = {Device Calibration Impacts Security of Quantum Key Distribution},
journal = {Physical Review Letters},
year = {2011},
volume = {107},
pages = {110501},
doi = {10.1103/PhysRevLett.107.110501}
}

@article{li_attacking_2011,
author = {Li, Hong-Wei and Wang, Shuang and Huang, Jian-Zhong and Chen, Wei and Yin, Zheng-Quan and Li, Fang-Yi and Zhou, Zhen and Liu, Dong and Zhang, Yong and Guo, Guang-Can and Bao, Wan-Su and Han, Zheng-Fu},
title = {Attacking a Practical Quantum-Key-Distribution System with Wavelength-Dependent Beam Splitter and Multiwavelength Sources},
journal = {Physical Review A},
year = {2011},
volume = {84},
number = {6},
pages = {062308},
doi = {10.1103/PhysRevA.84.062308}
}

@article{sun_passive_2011,
author = {Sun, Shi-Hai and Jiang, Min-Sheng and Liang, Lin-Mei},
title = {Passive Faraday-Mirror Attack in a Practical Two-Way Quantum-Key-Distribution System},
journal = {Physical Review A},
year = {2011},
volume = {83},
number = {6},
pages = {062331},
doi = {10.1103/PhysRevA.83.062331}
}

@article{fung_phase-remapping_2007,
author = {Fung, Chi-Hang Fred and Qi, Bing and Tamaki, Kiyoshi and Lo, Hoi-Kwong},
title = {Phase-Remapping Attack in Practical Quantum-Key-Distribution Systems},
journal = {Physical Review A},
year = {2007},
volume = {75},
number = {3},
pages = {032314},
doi = {10.1103/PhysRevA.75.032314}
}

@article{huang_laser_2019,
author = {Huang, Anqi and Navarrete, Alvaro and Sun, Shi-Hai and Chaiwongkhot, Poompat and Curty, Marcos and Makarov, Vadim},
title = {Laser Seeding Attack in Quantum Key Distribution},
journal = {Physical Review Applied},
year = {2019},
volume = {12},
pages = {064043},
doi = {10.1103/PhysRevApplied.12.064043}
}

@article{lo_mdi_2012,
  author = {Lo, Hoi-Kwong and Curty, Marcos and Qi, Bing},
  title = {Measurement-Device-Independent Quantum Key Distribution},
  journal = {Physical Review Letters},
  year = {2012},
  volume = {108},
  pages = {130503},
  doi = {10.1103/PhysRevLett.108.130503}
}

@article{curty_finite-key_2014,
author = {Curty, Marcos and Xu, Feihu and Cui, Wei and Lim, Charles Ci Wen and Tamaki, Kiyoshi and Lo, Hoi-Kwong},
title = {Finite-Key Analysis for Measurement-Device-Independent Quantum Key Distribution},
journal = {Nature Communications},
year = {2014},
volume = {5},
pages = {3732},
doi = {10.1038/ncomms4732}
}

@article{zeilinger_1999_foundational,
  author  = {Zeilinger, Anton},
  title   = {A Foundational Principle for Quantum Mechanics},
  journal = {Foundations of Physics},
  volume  = {29},
  number  = {4},
  pages   = {631--643},
  year    = {1999},
  doi     = {10.1023/A:1018820410908}
}

@article{russell_bell_state_analyzer_mdi_qkd,
  author  = {Russell, M. B. and Mailloux, L. O. and Hodson, D. D. and Grimaila, M. R.},
  title   = {A Bell State Analyzer Model for Measurement Device Independent Quantum Key Distribution},
  journal = {Proceedings of the International Conference on Scientific Computing (CSC'17)},
  year    = {2017},
  note    = {University/AFIT technical conference paper}
}

@article{chen_space_ground_2021,
  title={An integrated space-to-ground quantum communication network over 4,600 kilometres},
  author={Chen, Yu-Ao and Pan, Jian-Wei and others},
  journal={Nature},
  volume={590},
  pages={214--219},
  year={2021},
  doi={10.1038/s41586-020-03093-8}
}

\end{document}